\documentclass[11pt]{article}
\usepackage{epstopdf}
\usepackage[a4paper,hmarginratio=1:1,vmarginratio=2:3,totalwidth=15.2cm,totalheight=22.7cm]{geometry}
\usepackage{bm,epstopdf,epsfig,amsmath,amssymb,amsfonts,colordvi,latexsym,
comment,cancel,verbatim,
slashed}
\usepackage{epsfig,bbm}
\usepackage{graphicx,graphics}
\usepackage[font=md,captionskip=8pt]{subfig}
\usepackage[usenames,dvipsnames]{color}
\usepackage[noadjust]{cite}
\usepackage{xcolor}  
\usepackage{setspace}
\newcommand{\mm}{M}
\newcommand{\sg}{\sqrt{g}}    
\newcommand{\sge}{\sqrt{\hat g}}
\newcommand{\sgh}{\hat g}

\newcommand{\w}{\omega}
\usepackage[utf8]{inputenc}
\allowdisplaybreaks

\newcommand{\cK}{{\cal K}}
\newcommand{\cL}{{\cal L}}

\newcommand{\cO}{{\cal O}}

\newcommand{\cV}{{\mathcal V}}
\newcommand{\cZ}{{\cal Z}}

\newcommand{\ra}{\rightarrow}

\newcommand{\be}{\begin{equation}}
\newcommand{\ee}{\end{equation}}
\newcommand{\bea}{\begin{eqnarray}}
\newcommand{\eea}{\end{eqnarray}}

\newcommand{\baa}{\begin{array}}
\newcommand{\eaa}{\end{array}}

\long\def\symbolfootnote[#1]#2{\begingroup
\def\thefootnote{\fnsymbol{footnote}}\footnote[#1]{#2}\endgroup}
\begin{document} 
\begin{flushright}
\end{flushright}
\bigskip\medskip
\thispagestyle{empty}
\vspace{2.5cm}
\begin{center}

{\Large {\bf Spontaneous breaking of Weyl quadratic gravity}

\bigskip

{\bf  to Einstein action  and Higgs potential}}

\vspace{1.cm}

 {\bf D. M. Ghilencea} \symbolfootnote[1]{E-mail: dumitru.ghilencea@cern.ch}

\bigskip
{\small Department of Theoretical Physics, National Institute of Physics
 
and Nuclear Engineering, Bucharest\, 077125, Romania}
\end{center}

\bigskip
\begin{abstract}
\noindent
We consider the (gauged)  Weyl gravity action, quadratic in the scalar 
curvature ($\tilde R$) and in the Weyl tensor  ($\tilde C_{\mu\nu\rho\sigma}$) 
of the  Weyl conformal geometry.  In the {\it absence}  of  matter fields, this action 
has spontaneous breaking in which the Weyl gauge  field $\w_\mu$ becomes massive
(mass $m_\omega\!\sim$ Planck scale)  after   ``eating''  the dilaton 
in  the $\tilde R^2$ term, in a  Stueckelberg mechanism.
As a result, one recovers the Einstein-Hilbert action with a positive cosmological constant and
the Proca action for the massive Weyl gauge field~$\w_\mu$. Below $m_\w$ this field 
decouples and Weyl geometry  becomes Riemannian. The Einstein-Hilbert   action  is then
 just  a ``low-energy'' limit of  Weyl quadratic gravity which thus avoids 
 its  previous, long-held criticisms.
In the presence of matter  scalar field $\phi_1$ (Higgs-like),  with couplings allowed
by  Weyl gauge symmetry, after its spontaneous breaking one obtains in addition, at low scales,
a  Higgs potential with spontaneous electroweak symmetry breaking.
This is induced by the   non-minimal coupling $\xi_1\phi_1^2 \tilde R$ to Weyl geometry,
 with Higgs mass $\propto\!\xi_1/\xi_0$  ($\xi_0$ is the coefficient of the 
$\tilde R^2$ term).  In realistic models  $\xi_1$ must be classically tuned 
 $\xi_1\!\ll\!\xi_0$. We comment on the quantum stability of this value.
\end{abstract}

\newpage

\section{Motivation}

Introduced a century ago, Weyl's conformal geometry \cite{Weyl,Scholz,Dirac} is a 
scalar-vector-tensor theory of gravity that  is a generalization of
 Brans-Dicke-Jordan theory of gravity \cite{BDJ}.
The latter can be  recovered in the limit of vanishing Weyl gauge field.  
Weyl's conformal geometry was applied to model building beyond SM long ago
 \cite{Smolin,Cheng}, with   recent renewed interest 
\cite{JW,Quiros,Moffat1,Oh,Moffat2,ghilen, Heisenberg,Jimenez,P,Hill,Scholz2,Tann}.

In the {\it absence} of matter and up to  topological terms \cite{Tann},
the Weyl gravity action  is  the sum of two {\it quadratic}  terms:
the square of the Weyl-geometry scalar curvature term, $\tilde R^2$,  and the square of 
the Weyl-tensor: $\tilde C^2\equiv \tilde C_{\mu\nu\rho\sigma} \tilde C^{\mu\nu\rho\sigma}$.  
These generalise their counterparts $R^2$ and $C^2$ of the Riemannian geometry,
to  include effects due to the  Weyl gauge field $\w_\mu$. 
Indeed,  in Weyl geometry $\nabla_\mu g_{\nu\sigma}\propto \w_\mu g_{\rho\sigma}$,  
so the  Weyl connection  coefficients $\tilde \Gamma_{\mu\nu}^\rho$ are not determined  
by the metric alone (as in the Riemannian case), but also depend on $\w_\mu$.  
Then $\tilde R^2$  (or $\tilde C^2$) can be expressed in terms of their 
Riemannian counterparts $R^2$ ($C^2$)  plus a function of $\w_\mu$, and become
 equal if $\w_\mu=0$ (decoupled); in this limit, Weyl geometry becomes  Riemannian. 
We consider only a torsion-free Weyl geometry.

In Section~\ref{2} we show that in the {\it absence} of matter fields,
 Weyl quadratic gravity  has spontaneous breaking via a  Stueckelberg mechanism \cite{ES}
in which the Weyl gauge  field $\w_\mu$ becomes massive 
 after   ``eating''  the  Goldstone (dilaton)  field ($\phi_0$); 
$\phi_0$  is ``extracted'' from the $\tilde R^2$ term which propagates an 
extra scalar field.  Below the mass $m_\w$ of the Weyl gauge field,
$m_{\w}\!\sim$ Planck scale, this field decouples,
 Weyl geometry becomes Riemannian and  we recover the Einstein-Hilbert 
action and a positive cosmological constant.
So Einstein-Hilbert   action  is  a ``low-energy'' limit of 
 Weyl quadratic gravity which thus avoids previous long-held criticisms against it, 
see e.g.\cite{Scholz}. No additional matter field (scalar, etc) is required.

The Stueckelberg mechanism of breaking  is known in local Weyl-invariant models 
but these contain, however, {\it additional} (matter) scalars and are  {\it linear} 
(rather than quadratic) in the scalar curvature $\tilde R$ 
 of Weyl geometry \cite{Smolin,Cheng} (see also recent \cite{ghilen}).
In these models the Einstein action  follows from the presence
of a scalar matter field $\phi$  with  coupling $\phi^2\tilde R$,
 absent in our  quadratic Weyl action without matter.
To our knowledge, the breaking we study in the ``pure'' geometric case of   
$\tilde R^2$ and $\tilde C^2$-terms only, was not yet  discussed.
Our goal is to show that  this symmetry breaking mechanism is still at work
in this case.

In Section~\ref{3} we also consider the presence 
 of matter fields in addition to Weyl quadratic action.
We study the case of the  SM Higgs field ($\phi_1$) with all dimension-four 
couplings allowed by  Weyl gauge symmetry. 
We show that the Stueckelberg mechanism is still present, with
 an additional benefit: the Higgs potential  
has spontaneous electroweak symmetry breaking. This follows from a 
Weyl-invariant non-minimal coupling  $\xi_1 \phi_1^2 \tilde R$;
the higgs mass  becomes $\propto \xi_1/\xi_0$ (Planck units),
where $\xi_0$ is the coefficient of the $\tilde R^2$ term. 

There is also an ultraviolet (UV) motivation to study Weyl gauge symmetry: this
can play a role  in early cosmology
when  effective field theory at short distances becomes nearly conformal. Then models with
Weyl (gauged) symmetry
 \cite{Smolin,Cheng,JW,Quiros,Moffat1,Oh,Moffat2,ghilen, Heisenberg,Jimenez,Hill,P,Scholz2,Tann},
 conformal or global scale symmetry
\cite{Turok,tH0,TH,tH1,S1,Armillis,higgsdilaton,Monin,D1,D2,K1,R1,FRH1,FRH2,CW-BL,lebedev,Lalak,
Strumia}  provide an interesting setup for UV  model  building.
It is possible that Weyl quadratic  gravity  be a  renormalizable theory,
similar to (Riemannian) $R^2$ gravity \cite{KS}.
Thus Weyl gauge symmetry is interesting for studying the SM in 
the presence of gravity, with all scales  generated spontaneously.
This symmetry  may even be respected at the quantum level \cite{Englert,S1,Armillis,higgsdilaton,
Monin,D1,D2}. This opens the possibility to solve  the hierarchy problem
by endowing the SM with spontaneously broken Weyl gauge symmetry \cite{Moffat1,Oh,ghilen,Tann}.

\section{Spontaneous breaking of Weyl gauge symmetry}\label{2}

\subsection{From Weyl conformal geometry to a Riemannian description}

Let us  review some aspects of Weyl geometry  needed
 when discussing Weyl gauge invariance of the action.
 A (local) conformal  transformation of  the metric 
 and of a field $\phi$ is given by\footnote{
The conventions used are similar to those in \cite{R}, with  metric $(+,-,-,-)$.}
\bea
\label{ct}
g_{\mu\nu} &\ra&  g_{\mu\nu}^\prime =  e^{2\,\alpha(x)}\, g_{\mu\nu},
\nonumber\\ 
\phi &\ra& \phi^\prime   =  e^{-\alpha(x)\,\Delta}\,\phi, 
\eea
Then $g^{\mu\nu \prime}= e^{-2\alpha(x)} g^{\mu\nu}$ and
$\sqrt{g^\prime}\!=\! e^{4\alpha(x)}\! \sg$ with
$g\!=\!\vert\det g_{\mu\nu}\vert$.
$\Delta=1$ for a scalar field $\phi$  ($3/2$ for fermions). 
To transformation (\ref{ct}) one can associate that of a Weyl gauge field $\w_\mu$
which, due to (\ref{ct}), has geometric origin:
\bea\label{ct2}
\w_\mu\ra \w_\mu^\prime &=&\w_\mu-\frac{2}{q}\,\partial_\mu\alpha(x),
\eea

\noindent
Eqs.(\ref{ct}), (\ref{ct2}) define our Weyl (gauge) transformations;  
$q$ is the coupling to scalar field $\phi$,  with Weyl-covariant 
derivative $\tilde D_\mu\phi=(\partial_\mu-q/2\,\w_\mu)\phi$.
In Weyl geometry $\tilde\nabla_\mu g_{\alpha\beta}\!=-q\,\w_\mu\,g_{\alpha\beta}$,
with $\tilde\nabla=dx^\rho\tilde\nabla_\rho$ the Weyl covariant derivative computed 
with the  connection  $\tilde \Gamma_{\mu\nu}^\rho$, where
\medskip
\bea\label{tGamma}
\qquad
\quad\tilde\Gamma_{\mu\nu}^\rho=
\Gamma_{\mu\nu}^\rho+
\frac{q}{2}\,\Big[ \delta_\mu^\rho \,\w_\nu +\delta_\nu^\rho\,\w_\mu- g_{\mu\nu}\,\w^\rho\Big].
\eea

\medskip\noindent
 $\Gamma_{\mu\nu}^\rho$ is the  Riemannian (Levi-Civita) connection:
$\Gamma_{\mu\nu}^\rho=(1/2)\, g^{\rho\lambda}\, (\partial_\mu\, g_{\nu\lambda}+\partial_\nu\, g_{\mu\lambda}
-\partial_\lambda\, g_{\mu\nu})$.
Using eq.(\ref{tGamma}) one
verifies that  $(\tilde \nabla_\mu+ q\, \w_\mu)\,g_{\alpha\beta}=D_\mu\, g_{\alpha\beta}=0$,
with $D_\mu$ the Riemannian covariant derivative (computed with $\Gamma_{\mu\nu}^\rho$).
Under eqs.(\ref{ct}),  (\ref{ct2}),
$\tilde\Gamma_{\mu\nu}^\rho$ are invariant, which is not true for $\Gamma_{\mu\nu}^\rho$. 
The system is torsion-free (i.e. $\tilde\Gamma_{\mu\nu}^\rho=\tilde\Gamma_{\nu\mu}^\rho$) and
 $\tilde \Gamma_{\mu\nu}^\rho\ra  \Gamma_{\mu\nu}^\rho\, \, \text{if}\,\, \,\w_\mu\ra0$.

Then the  Riemann tensor in Weyl geometry, $\tilde R^\lambda_{\mu\nu\sigma}$,
 is  generated by the new $\tilde \Gamma_{\mu\nu}^\rho$  by a relation
 similar to that in the Riemannian geometry:
\medskip
\bea\label{tR}
\tilde R^\lambda_{\mu\nu\sigma}=
\partial_\nu \tilde\Gamma^\lambda_{\mu\sigma}
-\partial_\sigma\tilde\Gamma^\lambda_{\mu\nu}
+ \tilde\Gamma^\lambda_{\nu\rho}\,\tilde\Gamma^\rho_{\mu\sigma}
-\tilde\Gamma^\lambda_{\sigma\rho}\,\tilde\Gamma^\rho_{\mu\nu},
\eea

\medskip\noindent
and $\tilde R_{\mu\sigma\!}=\!\tilde R^\lambda_{\mu\lambda\sigma}$, 
$\tilde R\!=\!g^{\mu\nu} \tilde R_{\mu\nu}$.
With (\ref{tGamma}), (\ref{tR}),  $\tilde R$ is related to its Riemannian version ($R$):
\bea\label{tildeR}
\tilde R& =& 
R-3 \,q\,   D_\mu \w^\mu -\frac32 q^2 \, \w^\mu \w_\mu.
\eea

\medskip\noindent
Unlike $R$,  under  eqs.(\ref{ct}), (\ref{ct2}), $\tilde R$ transforms 
covariantly $\tilde R'=e^{-2\alpha(x)}\tilde R$, relevant below.

Finally, the Weyl tensor of Weyl geometry $\tilde C_{\mu\nu\rho\sigma}$,
like its Riemannian counterpart $C_{\mu\nu\rho\sigma}$, is a function
of associated   Riemann and Ricci tensors and scalar curvature\footnote{Explicitly 
$\tilde C_{\mu\nu\rho\sigma}=\tilde R_{\mu\nu\rho\sigma}
-(1/2) (g_{\mu\rho} \,\tilde R_{\nu\sigma} + g_{\nu\sigma} \tilde R_{\mu\rho} 
-g_{\mu\sigma} \tilde R_{\nu\rho} -
g_{\nu\rho} \tilde R_{\mu\sigma})
+(1/6) (g_{\mu\rho} \,g_{\nu\sigma} -g_{\mu\sigma} \,g_{\nu\rho}) \,\tilde R$.
A similar relation  exists for the Weyl tensor 
 $C_{\mu\nu\rho\sigma}$ of the Riemannian case, in terms of $R_{\mu\nu\rho\sigma}$,
$R_{\mu\nu}$ and $R$.
}. Then~\cite{Tann}
\bea\label{C}
\tilde C_{\mu\nu\rho\sigma}
=C_{\mu\nu\rho\sigma}
-\frac{q}{4}
\,(g_{\mu\rho}\,F_{\nu\sigma}+g_{\nu\sigma}\,F_{\mu\rho} -g_{\mu\sigma}\,F_{\nu\rho} -
g_{\nu\rho}\,F_{\mu\sigma})
+
\frac{q}{2} \,F_{\mu\nu}\,g_{\rho\sigma},
\eea

\medskip\noindent
where $F_{\mu\nu}$ is the field strength of the Weyl gauge field
$\w_\mu$; it is given by\footnote{In Weyl geometry
$\tilde F_{\mu\nu}=\tilde D_\mu\w_\nu-\tilde D_\nu\w_\mu$ and
 $\tilde D_\mu\w_\nu=\partial_\mu\w_\nu-\tilde \Gamma_{\mu\nu}^\rho\w_\rho$. 
In  Riemannian geometry
$F_{\mu\nu}=D_\mu\w_\nu-D_\nu\w_\mu$ and
 $D_\mu\w_\nu=\partial_\mu\w_\nu- \Gamma_{\mu\nu}^\rho\w_\rho$. Since 
$\tilde \Gamma_{\mu\nu}^\rho=\tilde \Gamma_{\nu\mu}^\rho$, 
$\Gamma_{\mu\nu}^\rho=\Gamma_{\nu\mu}^\rho$, 
then $\tilde F_{\mu\nu}=F_{\mu\nu}=\partial_\mu\w_\nu-\partial_\mu\w_\nu$.}
$F_{\mu\nu}=\partial_\mu\w_\nu-\partial_\nu\w_\mu$ and 
is  invariant  under transformations (\ref{ct}), (\ref{ct2}), 
and the same is true for $\sg\,\tilde C_{\mu\nu\rho\sigma}\tilde C^{\mu\nu\rho\sigma}$.

\subsection{Spontaneous breaking from  Weyl gauge symmetry to  Einstein gravity}

\noindent
In the absence of matter, the Weyl action has  two quadratic terms, with
\bea\label{totL}
L=L_1+L_2
\eea
Here
\bea\label{Rsquare}
L_1=\sg \, \,\frac{\xi_0}{4!} \,\,\tilde R^2, \qquad \xi_0>0.
\eea

\smallskip\noindent
Under transformations (\ref{ct}), (\ref{ct2}),   
$L_1$ is invariant. Another invariant under  (\ref{ct}), (\ref{ct2}) is
\medskip
\bea\label{Csq}
L_2 =\frac{\sg}{\eta}\,\,\tilde C_{\mu\nu\rho\sigma}\,\tilde C^{\mu\nu\rho\sigma}
=\frac{\sg}{\eta}\,\,\,\Big[ C_{\mu\nu\rho\sigma}\,  C^{\mu\nu\rho\sigma}+
\frac{3\,q^2}{2} \,F_{\mu\nu}\,F^{\mu\nu}\Big],
\eea

\medskip\noindent
where $\eta$ is the coupling in Weyl geometry and
 in the second step we used  eq.(\ref{C}). $L_2$ is decomposed 
into the sum of two terms of Riemannian geometry, each of them Weyl 
invariant. Note the presence of a kinetic term for $\w_\mu$, relevant below.
For this term to be canonically normalized in the presence of $L_2$, we 
must set\,\, $q^2=-\eta/6$ (so $\eta<0$).

In the absence of matter fields,  $L_{1,2}$ are the only independent
terms allowed by  Weyl gauge symmetry, up to  a topological term that is 
counterpart to the Riemannian Gauss-Bonnet term and that we do  not include here  
(see e.g. Appendix C in  \cite{Tann}).

Returning to $L_1$, $\tilde R^2$ has  higher derivatives, so
it propagates an additional scalar state, see e.g. \cite{DL,AK}. We extract from the $\tilde R^2$
term this (dynamical)  degree of freedom via a Lagrangian constraint;
this brings a term linear in $\tilde R$  and an additional  scalar  $\phi_0$. 
We have
\bea\label{eq}
L_1 = \sg\,\frac{\xi_0}{4!}\,\Big[ -2\, \phi_0^2\,\tilde R -\phi_0^4\,\Big]
\eea

\medskip\noindent
The equation of motion for $\phi_0$ gives $\phi_0^2=-\tilde R$.
Using this   back in (\ref{eq}),  one
recovers  $L_1$ of (\ref{Rsquare}), so Lagrangians (\ref{Rsquare}) and (\ref{eq}) are classically
equivalent.  Given its definition,  $\phi_0$ transforms just like  any matter  
scalar under eq.(\ref{ct}). Then   a (local) shift symmetry exists,
 as for a  Goldstone (dilaton) field:   $\ln\phi_0^2\ra \ln\phi_0^2-2\alpha(x)$.
This results from the Weyl invariance (\ref{ct}), (\ref{ct2}) of the term  $\phi_0^2\,\tilde R$;
note that its Riemannian counterpart ($\phi_0^2\, R$) does not have this symmetry,
 hence the importance of  this step  (obviously (\ref{eq}) is  
invariant under  (\ref{ct}), (\ref{ct2})).

Using eq.(\ref{tildeR}) to replace $\tilde R$ by its Riemannian version  
$R$ and an integration by parts, then
\bea
\label{L1-v2}
L_1=\sg\,\Big[ \,-\frac{\xi_0}{12}\,\phi_0^2\,R-\frac{q}{4}\,g^{\mu\nu}\,\w_\mu\,K_\nu 
+ \frac{q^2}{8}\,g^{\mu\nu}\,K\,\w_\mu\,\w_\nu
-\frac{\xi_0}{4!}\,\phi_0^4\,\Big]
+\text{total derivative.}
\eea

\noindent
where we introduced 
\bea\label{L1}
K_\nu=\partial_\nu K, \qquad
K=\xi_0\,\phi_0^2.
\eea

\medskip
Let us assume first that $L_1$ would be  the total Lagrangian of our model; 
its dependence on $\w_\mu$ 
is algebraic and this field can  be integrated out. Its equation of motion is
\bea
\w_\mu=\frac{1}{q}\,\partial_\mu\ln K
\eea

\medskip\noindent
Inserting this solution back in (\ref{L1-v2}) gives 
\medskip
\bea\label{ttt3}
L_{\rm eff}= \sg\,\Big\{ - \frac{\xi_0}{2}\,  \Big[\,\frac{1}{6} \phi_0^2\,R
 +g^{\mu\nu} \partial_\mu\phi_0\partial_\nu\phi_0\Big]
-\frac{\xi_0}{4!}\,\phi_0^4\,\Big\}.
\eea

\medskip\noindent
 $L_{\rm eff}$ is however uninteresting: although derived from 
(\ref{Rsquare}) and invariant under (\ref{ct}), 
it has a remaining ``fake'' conformal symmetry \cite{JP1}:
its associated current is vanishing. This follows 
the absence of a kinetic term for the Weyl  field
 which allowed its integration\footnote{
A second issue is that
with $\xi>0$,  $\phi_0$ becomes  ghost-like while if $\xi_0<0$
and with $\langle\phi_0\rangle\not=0$, 
Newton constant would be negative  (with our 
conventions). The Einstein term in our convention is
 $(-1/2)\,\sg M_p^2\,R$.}$^,$\footnote{
A third issue: a
 conformal transformation (\ref{ct}), $\alpha=-\ln(6 M_p/\xi_0\phi)$, ($M_p$: Planck scale) 
on (\ref{ttt3}) removes $\phi$  from spectrum and we recover the  Einstein action 
from $L_\text{eff}$, but then the number of degrees of freedom in Jordan vs Einstein frame
does  not match, so  ``something'' is missing;  see 
text  after eq.(\ref{L1DeltaL2}) for our solution.}.

This situation is  avoided  if $\w_\mu$ is dynamical, since then
the  current does not vanish anymore. A Weyl-invariant  kinetic term $\delta L_2$ 
for $\w_\mu$
\medskip
\bea\label{deltaL2}
\delta L_2=
-\frac{\sg}{4}\, g^{\mu\rho}\,g^{\nu\sigma}\, F_{\mu\nu}\, F_{\rho\sigma}
\eea

\medskip\noindent 
can be added 'by hand'; but there is no need to do so since  
$L_2$ of (\ref{Csq}) already contains $\delta L_2$!

To conclude, hereafter we shall consider that the 
defining action of our  model  is 
\bea\label{TotL}
L_1+\delta L_2
\eea
If one insists to also include the term $C^2\equiv C_{\mu\nu\rho\sigma} C^{\mu\nu\rho\sigma}$, 
then the action of our model is actually $L_1+ L_2$.
In both cases, the equation of motion for $\w_\mu$ gives
\medskip
\bea
\partial^\alpha (F_{\alpha\mu} \sg)+\frac{\sg}{2}\,\xi_0\,\phi_0 \Big[\partial_\mu -\frac{q}{2}
 \w_\mu\Big]\phi_0=0.
\eea

\medskip\noindent
By applying $\partial^\mu$ with  $F_{\alpha\mu}$  antisymmetric in $(\alpha, \mu)$,
we now find  a non-vanishing current  
\medskip
\bea\label{tildeK}
\tilde K_\mu\equiv\phi_0\,(\partial_\mu- q/2\,\w_\mu)\phi_0, 
\qquad \text{with}\qquad \partial^\mu(\sg\,\tilde K_\mu)=0,
\eea

\medskip\noindent
so $\tilde K_\mu$ is conserved.
Further, on the ground state, assuming $\phi_0(x)$=constant, it follows that
 $\partial^\mu(\sg\, \w_\mu)=0$, which is a condition similar to that for a
Proca (massive) gauge field, leaving  three degrees  of freedom for $\w_\mu$.
In fact,  in the case of  a Friedmann-Robertson-Walker (FRW) metric, 
$g_{\mu\nu}=(1,-a^2(t),-a^2(t),-a^2(t))$, with $\phi$
only $t-$dependent, the current conservation (in covariant form $D^\mu \tilde K_\mu=0$)  
leads naturally to $\phi_0$=constant \cite{FRH1}.

The other equations of motion, of $g^{\mu\nu}$ (after trace)  
and of $\phi_0$, derived from $L_1\!+\!\delta L_2$, are
\bea\label{eq18}
&&\frac{\xi_0}{12}\,\phi_0^2\,R+\frac{q}{4}\,g^{\mu\nu}\,\w_\mu\,K_\nu-
\frac{q^2}{8}\,K\,\w^\mu\,w_\mu+2 V+\frac{\xi_0}{4} \Box \phi_0^2=0
\qquad\eea
and
\bea\label{eq19}
&&\!\!\!\!\! 
-\frac{\xi_0}{12}\,\phi_0\,R+\frac{q^2}{8}\,\w_\mu\,\w^\mu\,\xi_0\,\phi_0-\frac12 \,V^\prime
+\frac{q}{4\,\sg}\,\xi_0\,\phi_0\,\partial_\mu (\sg\,\w^\mu)=0
\eea

\medskip\noindent
where we denoted $V\equiv(\xi_0/4!)\,\phi_0^4$. 
Eq.(\ref{eq19}) leads to $\langle\phi_0^2\rangle=-\langle\tilde R\rangle$ that we already know; 
thus the ground state has  $\tilde R$=constant (this is called Weyl gauge \cite{Scholz2}).
Further, when adding  eqs.(\ref{eq18}), (\ref{eq19}), with the 
last one multiplied by $\phi_0$, one finds  that on the ground state
\medskip
\bea
4\,V(\langle\phi_0\rangle)-\phi_0\,V^\prime(\langle\phi_0\rangle)=0
\eea

\medskip\noindent
The potential is thus a homogeneous function of fields, as expected (given the symmetry);
with our $V$, it is automatically respected;
$\langle\phi_0\rangle$ is thus a parameter, not fixed by theory; in
a Weyl-invariant  theory only (dimensionless) {\it ratios} of vev's can be fixed.

To  see how $\w_\mu$ becomes massive consider a 
conformal transformation  to Einstein frame 
\medskip
\bea\label{sc}
\sgh_{\mu\nu}=\Omega\,g_{\mu\nu}, 
\qquad
\Omega =\frac{\xi_0 \, \phi_0^2}{6\, \mm^2}
\eea

\medskip\noindent
$\mm$ is a  mass scale present for dimensional reasons\footnote{
One may avoid introducing $M$ if  $\sgh_{\mu\nu}$  is dimensionful \cite{TH}, which may 
be acceptable since  the metric is  a dynamical variable of kinetic term found in 
   $\tilde C_{\mu\nu\rho\sigma}^2  =C_{\mu\nu\rho\sigma}^2+...$ rather than a simple 
dimensionless constant.}; its role is discussed shortly. Then
\medskip
\bea\label{LL1}
\!
L_1\!=\!\sge\,\Big[
\frac{-1}{2}\,\mm^2\,\hat R \,
+
\frac{3\,\mm^2}{4\,\Omega^2}\,\hat g^{\mu\nu}\,
(\partial_\mu\Omega)(\partial_\nu\Omega)
+\frac{\hat g^{\mu\nu}}{\Omega}\Big(\frac{-q}{4} \w_\mu\,K_\nu
 +\frac{q^2}{8}  K\,\w_\mu\,\w_\nu
-\frac{\xi_0}{4!}\frac{\phi_0^4}{\Omega}\Big)
\Big].
\eea

\medskip\noindent
Thus the field $\phi_0$  is indeed dynamical,
 it has  a kinetic term.
Also note that $\delta L_2$ of (\ref{deltaL2}) 
is invariant under (\ref{sc}) since the metric part and $F_{\mu\nu}$ are invariant.  
Further, introduce
\bea\label{omega}
\w_\mu^\prime=\w_\mu-\frac{1}{q}\,\partial_\mu\ln K.
\eea

\medskip\noindent
Since  $K\!=\xi_0 \phi_0^2$ then 
$\phi_0$ is absorbed into $\w_\mu'$ in (\ref{omega}) 
where also $\partial_\mu\ln K=\partial_\mu\ln \Omega$.
Using (\ref{omega}),  replace $\w_\mu$ in $L_1$ and denote by $F^{\prime}_{\mu\nu}$ 
the field strength of  $\w^\prime_\mu$. Then the  total Lagrangian is 
\medskip
\be\label{L1DeltaL2}
L_1+\delta L_2 =\sge\,\Big[-\frac{1}{2}\, \mm^2\,\hat R -
\frac{3\, \mm^4}{2\,\xi_0}\, \Big]
+\sge\,\Big[-\frac{1}{4}\,\sgh^{\mu\rho}\,\sgh^{\nu\sigma}\, F^\prime_{\mu\nu}\, F^\prime_{\rho\sigma}
 +\frac{3\,q^2}{4}\,\mm^2\,\sgh^{\mu\nu}\w_\mu^\prime\,w_\nu^\prime \Big].
\ee

\medskip
It is important to note that there is no  kinetic term left 
for $\phi_0$, since it  was cancelled  by that generated
 when expressing $\w_\mu$-dependence in eq.(\ref{LL1}) 
in terms of  $\w_\mu^\prime$. The massless Weyl field  has become massive
after ``eating'' the Goldstone mode ($\phi_0$), via a  Stueckelberg mechanism,
{\it without} a corresponding Higgs mode  in the spectrum or a potential.
This mechanism essentially re-distributes the degrees of freedom in the action:
the initial massless $\w_\mu$ and the real scalar $\phi_0$ 
are converted into  a single massive Weyl field $\w_\mu$ with
three degrees of freedom (recall  $\partial^\mu(\sg\, \w_\mu)=0$);
so the number of degrees of freedom
 is indeed conserved when going
from the Jordan to the Einstein frame (as it should).

Having  checked this conservation,  eqs.(\ref{sc}), (\ref{omega})
may also be seen  as  a particular Weyl  transformation
eqs.(\ref{ct}),(\ref{ct2})  of  $2 \alpha(x)=\ln\Omega$, 
 to  the unitary gauge (``gauge fixing'') where the Goldstone  
$\phi_0$ is absent  (having been  ``eaten'' by now massive $\w_\mu'$) and
giving  $\hat\phi_0^{2}=\phi_0^2/\Omega=6 \mm^2/\xi_0=$constant.
The unitary gauge being non-renormalizable, one should not use 
(\ref{L1DeltaL2}) for loop  calculations, but use instead e.g. (\ref{L1-v2}), (\ref{deltaL2}).

In eq.(\ref{L1DeltaL2}) we obtained the Einstein-Hilbert action,
a positive cosmological constant and the  Proca action for a  massive 
Weyl  field\footnote{Our above result
is consistent with those  in \cite{DL,AK,YN} where it was shown that 'pure' 
$R^2$ gravity in the Riemannian geometry,  describes Einstein gravity plus a cosmological 
constant and a scalar (Goldstone) field. In our case the Goldstone mode  is  eaten
by the Weyl gauge field present in the $\tilde R^2$-term in Weyl geometry.} $\w_\mu'$; 
its mass is related to the  Planck scale ($M_p$) 
\bea
M_p^2=\mm^2,\qquad
m_{\w}^2=(3/2)\,q^2\,\mm^2.
\eea
%
The value of  $m_{\w}$  depends on the gauge coupling $q$
 in  $L_1+\delta L_2$,   see $L_1$ of (\ref{L1-v2})  and canonically 
normalised  $\delta L_2$ of (\ref{deltaL2}).
Below the scale $m_\w$ the Weyl  field decouples  and  Einstein gravity 
is obtained  as a ``low-energy'' effective theory limit.
 The scale $\mm$ introduced on dimensional grounds in (\ref{sc}),
remains undetermined by the theory.  From (\ref{L1-v2}),   
$\mm^2=\xi_0\langle\phi_0\rangle^2/6$ which  is equally undetermined in a theory  
invariant under (\ref{ct}), (\ref{ct2}),  as  discussed\footnote{In conformal theory
 only ratios of scales can be predicted in terms of dimensionless couplings.
If this symmetry is broken explicitly (by quantum corrections) 
dimensional transmutation can determine a field\,vev.}.

In the decoupling limit of the massive Weyl gauge field,
  Weyl geometry ``flows'' into a Riemannian geometry.
This may also  be seen dynamically from the conserved current in (\ref{tildeK}) 
which for  a  FRW metric  is driving $\phi_0$ to a constant value \cite{FRH1}
 (in this case $\propto M$). Below $m_\w$ the Weyl gauge field 
is absent, so  Weyl connection becomes that of the  Riemannian geometry, 
 $\tilde \Gamma_{\mu\nu}^\rho\!=\!\Gamma_{\mu\nu}^\rho$.
As a result, long-held  criticisms of Weyl quadratic gravity without matter
 \cite{Scholz,Scholz2} are avoided: the change of  the norm  
of a vector under parallel transport on
a closed curve or the change of the atomic spectral lines spacing
under Weyl transformation are effects  strongly suppressed by a very
high mass scale of Weyl gauge field,  $m_{\w}\propto M_p$.

So far our analysis was based on the Lagrangian  $L_1+\delta L_2$. Considering instead the 
Lagrangian $L=L_1+L_2$,
one has to include the remaining term in the rhs of (\ref{Csq}),  i.e.  
$C^2\equiv C_{\mu\nu\rho\sigma} C^{\mu\nu\rho\sigma}$; this is immediate, 
since this  is invariant under  (\ref{sc}), (\ref{omega}).
This term provides the kinetic term for the metric \cite{tH1}  and is needed 
at the quantum level to renormalize   divergences  like $k^4$. However,  
in this case the coupling $q$ cannot be adjusted at will anymore, being
proportional to  $\eta$ (see text after (\ref{Csq})). 
Lowering  $q$ too much brings a too light  mass $m_g^2\sim\eta\, M^2$ of the spin-two
ghost of the $C^2$ term, together with its instability.

It is interesting that Lagrangian (\ref{Rsquare}), (\ref{Csq}) 
dictated by Weyl geometry (no matter) is so rich in  structure, encoding 
 Stueckelberg mechanism, dilaton $\phi_0$,  Einstein action, Proca action 
for massive $\w_\mu$, a positive cosmological constant and fields kinetic terms
and  interactions.

\section{Adding matter fields} 
\label{3}

\subsection{Spontaneous breaking of Weyl gauge symmetry}

Let us now consider the SM scalar sector in addition to Weyl quadratic gravity action, 
with all dimension-four couplings allowed by the Weyl gauge symmetry. 
Note that the SM fermions do not have couplings to the Weyl gauge field
 (in the absence of torsion)
 \cite{Moffat1,ghilen}. 
We  consider  the  SM Higgs field and denote by $\phi_1$ its neutral 
component. Then the Lagrangian we study, invariant  under (\ref{ct}), (\ref{ct2}),
and written in a Weyl geometry language, is
\medskip
\bea\label{ll1}
\cL=\big(L_1+\delta L_2\big)
+ \sg\,\Big[
- \frac{1}{12} \,\xi_1\phi_1^2\,\tilde R
+\frac12 \,g^{\mu\nu}\,\tilde D_\mu\phi_1 \tilde D_\nu \phi_1 - V_1\Big]
\eea

\medskip\noindent
with Weyl-covariant derivative $\tilde D_\mu\phi_1\!=\!(\partial_\mu-q/2\,\w_\mu)\phi_1$ and 
$L_1$ and $\delta L_2$ of eqs.(\ref{Rsquare}),(\ref{deltaL2})
\medskip
\bea\label{fff}
L_1+\delta L_2=\sg\,\frac{\xi_0}{4!}\,\tilde R^2 - \frac{\sg}{4}\,g^{\mu\rho} g^{\nu\sigma}
F_{\mu\nu} F_{\rho\sigma}.
\eea

\medskip
The Weyl tensor squared term $C_{\mu\nu\rho\sigma} C^{\mu\nu\rho\sigma}$ in $L_2$ of (\ref{Csq})
 is not included in $\cL$, but since it is not affecting the transformations
below, it can easily be added (replace $\delta L_2 \ra L_2$). Also $F_{\mu\nu}$ is the same
in both Riemann and Weyl geometry (in the absence of torsion, as here).

The only possible form of the Higgs 
potential $V_1$ consistent with the symmetry, is
\bea
V_1=\frac{\lambda_1}{4!}\,\phi_1^4.
\eea
Using (\ref{eq}) and (\ref{tildeR}), one finds, following steps similar to the previous 
section\footnote{
In $V_1$ there is no classical coupling 
of $\phi_1$  to the dilaton ($\phi_0$) hidden  in Weyl's quadratic action (\ref{fff}) since
$\phi_0$ is an intrinsic part of our $\tilde R^2$ term from which is
extracted by ``linearisation'' of $\tilde R^2$, eq.(\ref{eq}).
Adding to $V_1$ a term 
$(\lambda_m/12)\,\phi_0^2\phi_1^2$ is also redundant, since together with
(\ref{eq}),  integrating out $\phi_0$  simply restores the original $\tilde R^2$ term 
after a  redefinition of initial couplings of $\phi_1$:
$\lambda_1\!\ra\! \lambda_1\!+\!\lambda_m^2/\xi_0$,\,\, $\xi_1\!\ra\! \xi_1\!+\!\lambda_m$.
Also,  adding a term  $\tilde\xi\phi_0^2 \tilde R$ to  (\ref{fff}) 
would introduce  an  extra (dynamical) degree of freedom beyond the dilaton
in $\tilde R^2$ term!}
\medskip
\be\label{ll2}
\cL=\delta L_2+
\sg\, \Big[
-\frac{1}{12} \xi_a\phi_a^2\,R
-\frac{q}{4}g^{\mu\nu}\w_\mu\partial_\nu \cK
+\frac{q^2}{8}\,g^{\mu\nu}\,\cK\,\w_\mu \,\w_\nu +
\frac12 \,g^{\mu\nu}\partial_\mu\phi_1 \partial_\nu\phi_1
-\cV(\phi_0,\phi_1)\Big]
\ee

\medskip\noindent
up to a total derivative, with
\smallskip
\bea
\cV&=&\frac{\xi_0}{4!}\,\phi_0^4+\frac{\lambda_1}{4!}\,\phi_1^4,
\\
\cK_\mu&=&\partial_\mu\,\cK, \qquad 
\cK=\xi_a\phi_a^2+\phi_1^2,\qquad \text{sum over}\,\, a=0,1.
\eea

\medskip\noindent
In (\ref{ll2}) a sum over the repeated index ``a'' is understood, with $a=0,1$. 
The generalisation of this action to more matter (scalar) fields is immediate.

We perform  a conformal transformation to the Einstein frame, to
 a new metric $\sgh_{\mu\nu}$ 
\bea\label{toE}
\sgh_{\mu\nu}=\Omega \,g_{\mu\nu},\qquad \Omega=\frac{1}{6 \mm^2}\, (\xi_0\phi_0^2+\xi_1\phi_1^2)
\eea
One has
\medskip
\bea\label{cL1}
\cL&=&\sge\,
\Big\{
-\frac12\, \mm^2\,\hat R
+\frac34 \,\sgh^{\mu\nu}\,\mm^2\,\partial_\mu\ln\Omega\,\partial_\nu\ln\Omega
\nonumber\\
&+&\frac{\sgh^{\mu\nu}}{\Omega} \Big[ 
 \frac12\,\partial_\mu\phi_1\,\partial_\nu\phi_1
-\frac{q}{4} \w_\mu\,\partial_\nu\,\cK
+\frac{q^2}{8}\,\,\cK\,\w_\mu\,\w^{\mu}\Big]
-\frac14 \,\sgh^{\mu\rho}\sgh^{\nu\sigma}\,F_{\mu\nu}\,F_{\rho\sigma}
-\frac{\cV}{\Omega^2} 
\Big\}
\eea

\medskip\noindent
Further, introduce
\bea\label{www}
\w_\mu^\prime=\w_\mu-\frac{1}{q}\,\partial_\mu\,\ln\cK.
\eea

\medskip\noindent
The kinetic terms  in $\cL$ for  $\phi_0$ and $\phi_1$ (hereafter $\cL_{k.t.}$) become,
after using eqs.(\ref{toE}), (\ref{www}) 
\medskip
\bea
\cL_{k.t.}\!&=&\!\sge\,\sgh^{\mu\nu}\,
\frac{1}{8\,\Omega}\,
\Big[\frac{6 \mm^2}{\Omega} (\partial_\mu\Omega)\,(\partial_\nu\Omega)
-\frac{1}{\cK} (\partial_\mu\cK)\,(\partial_\nu\cK)
+ 4 \,(\partial_\mu\phi_1)(\partial_\nu\phi_1)\Big]\quad
\\
&=&
\frac{3 \mm^2}{4} \frac{\sge\,\sgh^{\mu\nu}}{1+\cZ} (\partial_\mu\ln \cZ)(\partial_\nu \ln \cZ),
\quad \text{with}\quad
\cZ\equiv \xi_0\,\frac{\phi_0^2}{\phi_1^2}+\xi_1.
\label{k.t.}
\eea

\medskip\noindent
We see there is only one kinetic term left in the action, for the  new variable $\cZ$
which is a combination of initial $\phi_0$, $\phi_1$. 
One could  introduce polar coordinates fields $(\rho,\theta)$,
such as  $\phi_0=(1/\sqrt \xi_0)\, \rho\sin\theta$ and
$\phi_1=(1/\sqrt{1+\xi_1})\,\rho\cos\theta$. In such basis $\cZ$ is an ``angular'' 
variable field while $\cK$ entering in (\ref{www}) becomes 
$\cK=\rho^2$ and is the ``radial'' direction in field space.

After transformation (\ref{www}) the  terms\footnote{
These are the terms in (\ref{cL1})  proportional to $\mathcal V/\Omega^2$ and $\cK/\Omega$  
and clearly depend only of  the ratio $\phi_0/\phi_1$.}
in (\ref{cL1}) other than $\cL_{k.t.}$ also
depend only on the ratio $\phi_0/\phi_1\!\sim\! \cZ$ and
not on the radial direction field!
To anticipate, this is explained by $\w_\mu^\prime$ that must have ``eaten'' the 
radial direction in (\ref{www}) (Stueckelberg mechanism), see later.

We  bring to canonical form the kinetic term  $\cL_{k.t.}$ 
by replacing 
\medskip
\bea\label{lasts}
\frac{1}{\cZ}={\sinh^2 \frac{h}{M\sqrt 6}},
\eea

\medskip\noindent
where  $h$ is our (neutral) Higgs field.
This gives that  $\cL_{k.t.}=(1/2)\,\sqrt{\hat g}\,\hat g^{\mu\nu}(\partial_\mu h)(\partial_\nu h)$.

From  $\cL$ of (\ref{cL1}), using (\ref{k.t.}) and notations  (\ref{www}), (\ref{lasts}),
we obtain our final Lagrangian
\medskip
\bea\label{cL4}
\cL=\!\sge
\Big[
-\frac{\mm^2}{2} \hat R
+\frac{\sgh^{\mu\nu}}{2} (\partial_\mu h)(\partial_\nu h)
-\frac14 \,\sgh^{\mu\rho}\sgh^{\nu\sigma}\, F^\prime_{\mu\nu} F^\prime_{\rho\sigma}
+\frac{m_\w^2(h)}{2}\w_\mu^\prime\,\w^{\prime \mu}
-\hat\cV(h)\Big],
\eea

\medskip\noindent
where $F_{\mu\nu}^\prime$ is the field strength of $\w_\mu^\prime$ and
\medskip
\bea
m^2_\w(h)&=&\frac{q^2 \cK}{4\,\Omega}=
\frac{3}{2} \,\mm^2\,q^2\,\cosh^2 \frac{h}{M\sqrt 6},
\label{ex1}
\\
\hat{\mathcal V}(h)&=&\frac{\mathcal V}{\Omega^2}
=
\frac32 \frac{M^4}{\xi_0}
\Big[ 1 - 2 \,\xi_1\, \sinh^2 \frac{h}{M\sqrt 6}
+\big(\lambda_1\,\xi_0 +\xi_1^2\big)\,\sinh^4\frac{h}{M\sqrt 6}\Big].
\label{ex2}
\eea

\medskip\noindent
For small higgs field values $h\ll M$ one has
\medskip
\bea\label{e1}
 m^2_\w( h)&=&\frac{3}{2} \,\mm^2\,q^2\,\Big[
1+\frac{h^2}{6 \,\mm^2}+ \frac{h^4}{108\,\mm^4}+\cO(h^6/\mm^6)\Big]
\\[8pt]
\hat\cV(h)&=&\frac{3\,\mm^4}{2\,\xi_0}-\frac{\xi_1\,\mm^2}{2\,\,\xi_0} \, h^2+
\frac{1}{4!}\,\Big[\lambda_1+\frac{\xi_1\,(3\xi_1-2)}{3\,\xi_0}\Big]\, h^4+\cO( h^6/\mm^6)
\label{e2}
\eea

\medskip
The first term on the rhs of (\ref{e1})  is the mass of the Weyl gauge field $\w_\mu^\prime$, up 
to additional corrections of order $\cO(\langle h\rangle^2\!/M^2)$ 
due to the Higgs mechanism itself, if $\langle h\rangle\not\!=\!0$  (see below). 
 Therefore, following eq.(\ref{www}) the   ``radial'' degree of freedom in the field space
(dilaton)  was ``eaten'' by the gauge field $\w^\prime_\mu$ which has become massive, via
the Stueckelberg mechanism. The number of degrees of freedom is conserved: initially we  had
2 real scalars ($\phi_0, \phi_1$) and massless $\w_\mu$ which were re-arranged into
one real scalar $h$ and a massive gauge field $\w^\prime_\mu$.

$\cL$ in eq.(\ref{cL4})  includes
 the Einstein action with a positive cosmological constant,
a massive Weyl gauge field   and a potential for the Higgs field $h$. 
For a vanishing non-minimal coupling  $\xi_1\!=\!0$, see our starting  
$\cL$ in (\ref{ll1}),  one recovers the
 initial potential for $\phi_1$ plus a cosmological constant similar to
 that in Weyl quadratic action without matter, eq.(\ref{L1DeltaL2}).

 Eqs.(\ref{cL4}) to  (\ref{ex2}) give the Higgs sector for the SM enlarged  with
Weyl gauge symmetry and can be used for further investigations of this symmetry.
These equations bring no restrictions at the classical level
 for the  value of $h$  relative to (otherwise arbitrary) $M$.

\subsection{Electroweak symmetry breaking}

From (\ref{ex2}), (\ref{e2}), 
 we see that a non-minimal coupling $\xi_1\tilde R^2 \phi_1^2$,\,($\xi_1\!>\!0$)
induced a negative quadratic term  in the potential  and spontaneous
electroweak symmetry breaking, with
\medskip
\bea
m_h^2=\frac{\xi_1}{\xi_0}\,\mm^2.
\eea

\medskip\noindent
A hierarchy $m_h^2\!\ll\! \mm^2$ can be arranged by a classical tuning
$\xi_1\ll \xi_0$ to an ultraweak value of $\xi_1$. 
This is a  gravitational higgs mechanism (which forbids the presence
of TeV-scale squarks!)\footnote{The quartic term in (\ref{e2}) remains positive as long as
 $3\lambda_1 \xi_0\!>\! (2-3\xi_1)\xi_1$\, ($\lambda_1>0$, $\xi_1>0$)
which is easily satisfied for small enough $\xi_1\ll \xi_0$ required for small Higgs mass. 
 Also note from eq.(\ref{ex2}) 
 that at large $h$, the quartic term receives overall positive corrections from 
the gravitational effects, proportional to $\xi_1^2/\xi_0$.}.

Regarding  the scale $M$, in the limit  $h\ll M$, can be identified
with a constant value of the  dilaton  $\langle\phi_0\rangle$ (in this limit
$\langle\phi_0\rangle \gg\langle\phi_1\rangle$);
like $M$, $\langle\phi_0\rangle$ is a parameter not fixed by the theory
and must be tuned to the actual Planck scale value (as mentioned, only ratios of 
scales can be determined in a Weyl invariant theory). Finally, 
the vacuum energy is still positive, dominated by the dilaton contribution
$\hat\cV_{\text{min}}=3\,\mm^4/(2\xi_0)[1-\xi_1^2/(\xi_0\lambda_1)+...]$.

A study of the quantum corrections to the Higgs mass is beyond the goal of this paper. 
However, we stress here the role  Weyl gauge symmetry may play at the quantum level.
Note that  classically only the Higgs sector of the SM couples to the Weyl field $\w_\mu$
\cite{Moffat1,ghilen}.  Using a Weyl-invariant
regularization \cite{Englert} one could answer whether Weyl gauge symmetry can
 protect $\xi_1$ and thus the Higgs mass  $m_h\propto \xi_1$  against large quantum
 corrections\footnote{without  additional fine tuning beyond the classical one 
mentioned above.}.
Note that each of the terms in our $\cL$ of (\ref{ll1}), (\ref{fff}) is separately
 Weyl gauge invariant, see eqs.(\ref{ct}), (\ref{ct2}). We expect that this symmetry bring 
some ``protection'' for the ultraviolet (UV) behaviour of this theory. In
 particular we expect a  better UV behaviour than in   Riemannian gravity 
e.g. \cite{Strumia} where no similar local symmetry (Weyl, conformal) exists.
This motivates a quantum analysis of Weyl quadratic gravity.

In the light of the results in  \cite{KS} for renormalizability,  
one would  expect  a Weyl quadratic theory given by 
$L_1+L_2$ of (\ref{Rsquare}), (\ref{Csq})  be  renormalizable. 
One must however pay attention to the  analytical continuation 
from Minkowski to the Euclidean space which is non-trivial 
in the presence of higher derivative terms \cite{IA}.

\section{Conclusions}

We considered the  general action of Weyl gravity  in the absence of matter, 
which  is  the sum of two terms {\it quadratic} in the curvature scalar ($\tilde R$)  and in 
the  Weyl tensor ($\tilde C_{\mu\nu\rho\sigma}$) of the Weyl conformal geometry,  
then studied its spontaneous breaking. We also studied the effect of coupling  Weyl 
(quadratic) gravity to a Higgs-like matter sector.
 
In the absence of  matter fields,  the Weyl gauge field $\w_\mu$ in the action
 becomes  massive, with  a mass  $m_\w\sim q\, M_p$ where  $q$ is the 
coupling and $M_p$ the Planck scale.  This happens  via a Stueckelberg mechanism 
which    is essentially a re-arrangement of the degrees of freedom (without a 
higgs vev or potential needed):
the field $\w_\mu$ ``eats'' a scalar degree of freedom (dilaton) 
 ``extracted''  from the  $\tilde R^2$ term. 
The necessary presence 
of a kinetic term for the Weyl gauge field originates from the term 
 $\tilde C_{\mu\nu\rho\sigma} \tilde C^{\mu\nu\rho\sigma}$  of Weyl geometry.
However, if this is not included in the initial Weyl quadratic action, the
 gauge kinetic  term   may be added on its own in Weyl geometry,  on symmetry arguments
only.

After the Stueckelberg mechanism one obtains the Einstein-Hilbert action,
a positive cosmological constant and Proca action for the massive Weyl gauge field
 (and a Riemannian Weyl-tensor-squared term,  if 
$\tilde C_{\mu\nu\rho\sigma}\tilde C^{\mu\nu\rho\sigma} $  was included initially).
No additional matter scalar field (Higgs, etc) is needed to this purpose.
Below the mass $m_\w$  the Weyl field  $\w_\mu$  decouples and Weyl 
geometry (connection)  becomes Riemannian. 
Therefore, the Einstein-Hilbert action is   a  ``low-energy'' effective theory 
limit of  Weyl quadratic  gravity (without matter). In this way
Weyl quadratic gravity  avoids  previous, long-held criticisms against~it. 

This result  has consequences for physics at high scale where Weyl gauge symmetry 
may be present  (inflation, black hole physics, conformal supergravity). 
During the  spontaneous breaking of this symmetry the number of degrees of freedom 
is indeed conserved, also when going  to the unitary gauge $\hat\phi_0$=constant
 (unlike in models invariant under ``usual'' Weyl symmetry, eq.(\ref{ct})).
It is  remarkable that the simple Weyl quadratic action  dictated by Weyl 
geometry alone is so rich  in structure, encoding  a Stueckelberg mechanism, 
the Einstein-Hilbert action, Proca action,  positive  cosmological constant,    
the dilaton, the metric and their interactions.  Further study of this symmetry should use
the Weyl geometry formulation which is easier (than the Riemannian one)
since then the  scalar curvature  transforms covariantly, so each operator respects 
this symmetry.  This   places on equal footing, in the Lagrangian,  Weyl gauge 
symmetry and  (internal) gauge symmetries.

In the presence of a scalar matter field $\phi_1$  (Higgs), the Weyl gauge symmetry  allows
 a non-minimal  coupling $\xi_1\phi_1^2\,\tilde R$, in addition to the mentioned Weyl quadratic 
action and to the matter action; the latter is that of the SM Higgs sector
with a  potential $\lambda_1\phi_1^4/4!$ (as the only one allowed by this symmetry). 
The Stueckelberg breaking mechanism 
is still at work and the Weyl gauge field is ``eating'' the dilaton 
(the radial direction in the field space of $\phi_0$, $\phi_1$) which subsequently
 disappears from the action. At the same time, in the Riemannian limit ($\w_\mu$ decoupled), 
the scalar potential
of the remaining Higgs degree of freedom acquires at low energy ($h\!\ll\! \mm$) 
 a negative quadratic term $\propto\! \xi_1/\xi_0$.
 This is  gravitationally-induced spontaneous electroweak symmetry breaking.
It is worth investigating further if the Higgs mass value
($\xi_1/\xi_0$ in Planck units) is stable  at the quantum level
 in  SM with spontaneously broken Weyl gauge symmetry.

\bigskip\bigskip\bigskip
\noindent
{\bf Note added:} After completing  this work we became aware of 
\cite{RP1} where the connection in a general  theory of gravity 
is shown to acquire  mass via a Higgs mechanism, while at low scales 
is ``frozen'' to the Levi-Civita connection. This is consistent with our result.

\bigskip\noindent
{\bf Acknowledgements:} We thank Mihai Visinescu (IFIN Bucharest) and Hyun Min Lee 
(Chung-Ang University, Seoul) for   discussions on Weyl geometry and its applications.

\end{document}